\PassOptionsToPackage{bookmarks=false}{hyperref}

\documentclass[conference]{IEEEtran}
\IEEEoverridecommandlockouts
\usepackage{cite}
\usepackage{xcolor}

\usepackage{amsmath,amsfonts, amssymb, amsthm, xcolor}
\usepackage{algorithmicx} 
\usepackage{algorithm}
\usepackage{algpseudocode}

\makeatletter
\@removefromreset{algorithm}{section}

\makeatother

\usepackage{array}
\usepackage[caption=false,font=normalsize,labelfont=sf,textfont=sf]{subfig}
\usepackage{textcomp}
\usepackage{stfloats}
\usepackage{verbatim}
\usepackage{graphicx}
\usepackage{tikz}
\usetikzlibrary{arrows.meta, positioning, decorations.pathreplacing}
\usepackage{pgfplots}
\pgfplotsset{compat=1.18} 
\usepgfplotslibrary{patchplots,fillbetween}
\usepackage{orcidlink}
\definecolor{orcidlogocol}{HTML}{A6CE39}

\usepackage[bookmarks=false]{hyperref}
\usepackage[numbers]{natbib}
\setcitestyle{aysep={}} 

\newcommand{\ba}{\mathbf{a}}

\newcommand{\bh}{\mathbf{h}}
\newcommand{\bx}{\mathbf{x}}

\newcommand{\bw}{\mathbf{w}}

\newcommand{\bp}{\mathbf{p}}
\newcommand{\bt}{\mathbf{t}}

\newcommand{\bff}{\mathbf{f}}
\newcommand{\bg}{\mathbf{g}}

\newcommand{\bX}{\mathbf{X}}
\newcommand{\bY}{\mathbf{Y}}

\definecolor{darkgreen}{RGB}{0,155,0}

\begin{document}

\title{{\fontsize{23.8}{26}\selectfont CKM-Assisted Physical-Layer Security for Resilience Against Unknown Eavesdropping Location}
\thanks{This work has been funded by the LOEWE Initiative, Hesse, Germany, within the emergenCITY Centre under Grant LOEWE/1/12/519/03/05.001(0016)/72.}
}

\author{\IEEEauthorblockN{Ladan Khaloopour~{\orcidlink{0000-0002-3008-2292}}, Matthias Hollick~{\orcidlink{0000-0002-9163-5989}}, and Vahid Jamali~{\orcidlink{0000-0003-3920-7415}}}
\IEEEauthorblockA{\textit{Technical University of Darmstadt, Darmstadt, Germany} 
}
}
\maketitle
\begin{abstract}
Channel Knowledge Map (CKM) is an emerging data-driven toolbox that captures our awareness of the wireless channel and enables efficient communication and resource allocation beyond the state of the art. In this work, we consider CKM for improving physical-layer security (PLS) in the presence of a passive eavesdropper (Eve), without making any assumptions about Eve's location or channel state information (CSI). We employ highly directional mmWave transmissions, with the confidential message jointly encoded across multiple beams. By exploiting CKM, we derive an algorithm for time and power allocation among the beams that maximizes the absolute secrecy rate under the worst-case scenario for Eve's location.
\end{abstract}


\section{Introduction}
In recent years, channel knowledge map (CKM)\footnote{CKM is also known by other names, e.g., radio frequency (RF) map, channel charting, time-domain channel prediction, etc. \cite{studer2018channel, jiang2019neural, duel2007fading}.} has received considerable attention as a new tool for improving communication, resource allocation, transmission robustness, and more \cite{studer2018channel, zeng2024tutorial}. CKM includes contextual information about the wireless channel, which can be collected from a variety of sources, including real-world data from previous transmissions, artificial intelligence-based channel simulators, and various sensors. In this paper, we investigate the benefits of CKM for physical-layer security (PLS).

Most works on PLS assume that the perfect or imperfect CSI of Eve, its location area, or the direction toward Eve 
are available  
\cite{ lin2017physical, valliappan2013antenna, xu2022robust}. However, these assumptions might not be feasible in practice for a passive uncooperative eavesdropper. 
PLS can be enhanced by using higher frequency bands, such as millimeter wave (mmWave) frequencies, where the leakage probability is reduced due to the highly directional beams. However, the line-of-sight (LoS) link remains insecure to passive eavesdropping.

In this paper, we exploit CKM to enable PLS at mmWave bands, even in scenarios where no knowledge of Eve's CSI or location is available. The confidential message is jointly encoded across multiple beams, which makes it impossible for the eavesdropper to recover the message unless it can decode all beams. This is physically unlikely, as the eavesdropper is present at only one (unknown) location. By leveraging CKM, we derive an algorithm for time and power allocation among the beams, that maximizes the absolute secrecy rate under the worst-case scenario for Eve's location, i.e., the strongest attack location of Eve. Unlike our previous works \cite{ishtiaq2023beamsec, ishtiaq2025twc}, which considered only time allocation and provided no analytical solution, here we derive analytical solutions for both time and power allocation.

\section{System Model and Preliminary} 
\subsection{System Model}
\begin{figure}[t]
\centering
\scalebox{1.0}{
\begin{tikzpicture}[
    block/.style={rectangle, draw, minimum height=0.7cm, minimum width=1.4cm, align=center},
    smallblock/.style={rectangle, draw, minimum height=0.7cm, minimum width=0.8cm, align=center},
    bigblock/.style={rectangle, draw, minimum height=0.7cm, minimum width=1.1cm, align=center},
    circleblock/.style={circle, draw, minimum size=0.9cm, align=center},
    sumcircle/.style={circle, draw, minimum size=0.4cm, font=\bfseries\small, inner sep=0pt},
    arrow/.style={-{Latex[width=1.5mm]}, thick},
    node distance=0.5cm and 0.8cm
]

\clip (-1.5,-1.2) rectangle (4.3,1.2); 

\node[smallblock] (tx) {$\bw$};
\node[bigblock, right=of tx] (channel) {$\bh_i$};
\node[sumcircle, right=of channel] (sum) {$+$};
\node[block, right=of sum] (rx) {$\bff$ or $\bg$};

\node[left=of tx] (input) {};
\draw[arrow] (input) -- (tx) node[midway, above] {$x$};
\draw[arrow] (tx) -- (channel) node[midway, above] {$\bx$};
\draw[arrow] (channel) -- (sum);
\draw[arrow] (sum) -- (rx) node[midway, above] {$y_i$};
\node[right=of rx] (input2) {};
\draw[arrow] (rx) -- (input2) node[midway, above] {$y$};
\node[above=of sum] (input3) {$z_i$};
\draw[arrow] (input3) -- (sum);

\node[below=0.3cm of tx] (z) {Tx};
\node[below=0.3cm of channel] (z) {Channel};
\node[below=0.45cm of sum] (z) {AWGN};
\node[below=0.3cm of rx] (z) {Rx or Eve};

\end{tikzpicture}
} 
\caption{General block diagram of the system model ($i\in\{r,e\}$).}
  \label{fig: system_model}
\end{figure}
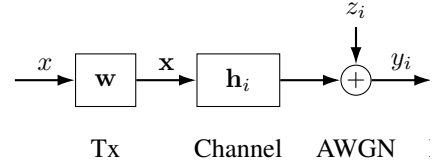

We consider a wireless communication system including a transmitter (Tx), a legitimate receiver (Rx), and a passive eavesdropper (Eve). Tx sends a confidential message to Rx, and Eve tries to intercept this message. 
Tx has $N_t$ antennas, and for simplicity, Rx and Eve have a single antenna, and Eve's location is unknown. The confidential message is denoted by $x \in \mathbb{C}$, and $\bw$ is the beamforming vector of the transmitter.
Therefore, the transmitted data symbol $\bx \in \mathbb{C}^{N_t}$ is given by
\begin{align}
\bx = \bw \, x.
\end{align}
We assume $\mathrm{E}\{|x|^2\} \leq P_\text{tx}$, where $\mathrm{E}\{\cdot\}$ is the expectation operator.
The received signal is shown~as 
\begin{align}
y_i = \bh_i^\text{H} \, \bx + z_i, \quad i\in\{r,e\},
\end{align}
where subscripts $r$ and $e$ denote Rx and Eve, respectively. $y_i \in \mathbb{C}$, and $z_i \sim \mathcal{C}\mathcal{N}(0,\sigma_i^2) \in \mathbb{C}$ is the additive white Gaussian noise (AWGN), 
$\bh_i\in \mathbb{C}^{N_t}$ is the channel matrix that models the multi-path propagation of the transmitted signal in the channel between Tx and node $i$.
The system model is shown in Fig.~\ref{fig: system_model}. 
The received signal to noise ratio (SNR) is 
\begin{align}
\gamma_{i} = \frac{P_\text{tx}}{\sigma_i^2}{\,|\bh_i^\text{H} \, \bw|^2}, \quad i\in\{r,e\}.
\end{align}
We assume that Tx knows the CSI of Rx, i.e., $\bh_r$, 
but does not know Eve's location $\bp_e$, its SNR $\gamma_{e}$, or channel $\bh_e$. 

\subsection{CKM} 
CKM provides information about how RF is distributed in the environment. This of course depends on the direction that Tx transmits, the environment itself, and the random objects. 
For rigorousness, we consider the following CKM. 

\textit{Definition}: Let ${\gamma}(\theta, \bp)$ be the expected power that is collected by Rx at location $\bp$ due to radiation by transmitter with power $P_\text{tx}$ along direction $\theta$, i.e., $\bw = \ba_\text{tx}(\theta)$, where 
$${\ba}_\text{tx}(\theta) = \frac{1}{\sqrt{N_t}}[1, e^{-j\frac{2\pi}{\lambda}d \sin(\theta)}, \cdots , e^{-j\frac{2\pi}{\lambda}(N_t-1)d \sin(\theta)}]^{\text{T}}.$$
Throughout this paper, CKM is referred to any (experimental or simulation-based) knowledge about  
random variable (RV) ${\gamma}(\theta, \bp), \forall \theta \in \Theta\doteq \{\theta_l|l=1,\cdots,L\} \text{ and } \forall \bp \in \hat{\mathcal{P}} \doteq \{\bp_j|j=1,\cdots,J\}$, where $\Theta$ and $\hat{\mathcal{P}}$ are the set of angle of departure (AoD) and Rx location for which the CKM is constructed.

\section{CKM-assisted Security Improvement} \label{sec:DT_assisted_security}

In this section, first we provide a concise overview of the physical layer security conditions and then introduce the secure coding scheme considered in this work.

\subsection{Physical Layer Security and Secure Coding}
The Tx aims to transmit the confidential message $\bX$ of length $n$ over the channel to the Rx in the presence of an Eve. Let us assume that Rx decodes the message as $\hat{\bX}$ and the Eve's observations are $\bY_e$. The communication between Tx and Rx is ensured to be reliable and secure if the two following conditions are satisfied \cite{yeh2022angularly, schaefer2015secure}:
\begin{subequations}
\begin{align}
    & \lim_{n \rightarrow \infty} \text{Pr}(\bX\neq \hat{\bX})=0, \\
    & \lim_{n \rightarrow \infty} \frac{1}{n} I(\bX;\bY_e)=0,
\end{align}
\end{subequations}
where $\text{Pr}(\cdot)$ and $I(\cdot;\cdot)$ show the probability and the mutual information operators, respectively. According to \cite{poor2017wireless,yeh2022angularly, schaefer2015secure}, 
the conditions of reliable and secure communication between Tx and Rx are satisfied if the transmission rate $R_{r}$ is bounded~as 
$$
R_{r} \leq [C_{r}-C_{e}]^+,
$$
where the right hand side of the above inequality shows the secrecy capacity. For Gaussian wiretap channel,  
we have
$$C_{i} \doteq \log_2(1+\gamma_{i}), \quad i\in\{r,e\}$$ 
as the channel capacity of node $i$, where $[q]^+ \doteq \max(q,0)$.

\subsection{Secure Coding Over Multi-beams}
We use the mmWave band for transmission, which has highly directional and narrow beams, and helps secure communication by reducing the signal footprint.   
However, when the LoS is used for signal transmission, the secrecy rate can be zero if Eve is located at a point on the LoS path between Tx and Rx (due to the stronger Eve channel).  
Here, we use multiple beams and follow the idea of jointly encoding the message over different beams proposed in \cite{yeh2022angularly}. 
Eve can intercept part of the message if it is located in the path of a beam. But, it cannot intercept all beams to decode the entire message. Thus, the message remains secure.

We denote the fractions of time and power allocated to path (or AoD) $\theta_l$ by $t_l$ and $p_l$, respectively. 
The node $i$ capacity for the $l$-th transmission is $$C_{i,l} = \log_2(1+\gamma_i(\theta_l, \bp)). $$
Since Tx does not know Eve's location and channel, we consider the secrecy capacity for the worst-case location $\bp_e$, and channel realization $\bh_e(\bp_e)$~as:
\begin{align} \label{eq:secrecy_capacity}
C_s = \min_{\substack{\bp_e \in {\mathcal{P}}} } \,\,\,\, \sum_{l=1}^{L} [C_{r,l}-\max_{\substack{\bh_e(\bp_e) \in \mathcal{H}(\bp_e)}}\,\, C_{e,l}]^+,
\end{align}
where ${\mathcal{P}}$ is the set of points we would take to ensure secrecy and $\mathcal{H}(\bp_e)$ is the set of fading realization that a potential Eve at location $\bp_e$ may experience.
In practice, the set ${\mathcal{P}}$ cannot be the entire environment since otherwise the secrecy rate will be strictly zero (e.g., at least an eavesdropper arbitrary close to the Rx can recover the message). 
Moreover, set $\mathcal{H}(\bp_e)$ can be in practice unknown. In this paper, we aim at approximating $C_s$ (and ${\mathcal{P}}$) 
by the use of CKM, which provides an approximation of $\mathcal{H}(\bp_e)$ (and ${\mathcal{P}}$), denoted by $\mathcal{H}^\text{CKM}(\bp_e)$ (and ${\hat{\mathcal{P}}}$).

\subsection{Problem Formulation}
Our objective is to maximize the secrecy capacity 
by jointly optimizing the time fractions $t_l$ and power $p_l$ allocated to beam $l,\,\,\forall l \in \{1, \cdots, L\}$. We define an estimated secrecy rate $\hat{C}_s$, which is obtained from the available CKM, i.e., using  $\mathcal{H}^\text{CKM}(\bp_e)$ instead of $\mathcal{H}(\bp_e)$ in \eqref{eq:secrecy_capacity}. Therefore, we have
\begin{align} \label{P1_opt_prob}
    \textbf{P1:} \,\,\,\, &\underset{p_l \ge 0, t_l \ge 0, \forall l}{\text{max}} \,\,\,\,  \min_{\forall \bp_j \in \hat{\mathcal{P}}} \,\,\,\, \sum_{l=1}^{L} t_l \Big[ \log_2 \big( \frac{1+p_l \alpha_l}{1+p_l \beta_{l, j}} \big)\Big]^+\quad  \nonumber \\
    &\text{s.t. } \quad\,\,\, \text{C}_1 : \sum_{l=1}^{L} t_l \leq 1, \quad  \text{C}_2 : \sum_{l=1}^{L} p_l t_l \leq P_\text{tx},   
\end{align} 
where $\alpha_l\doteq \frac{\gamma_r(\theta_l)}{P_\text{tx}}$ and $\beta_{l,j}\doteq \underset{\gamma(\theta_l,\bp_j)\in \mathcal{H}^\text{CKM}(\bp_j)}{\max}\frac{\gamma(\theta_l,\bp_j)}{P_\text{tx}} $ are the  normalized SNRs  at Rx and $\bp_j$, respectively, when beam $l$ is adopted. In problem \textbf{P1},  
the objective function ensures absolute secrecy regardless of Eve's location within $\bp_j \in \hat{\mathcal{P}}$ and constraints $\text{C}_1$ and $\text{C}_2$ enforce the total time and power budgets, respectively.

\subsection{Solution Development}
The optimization problem \textbf{P1} is not convex due to the operator $[\cdot]^+$ and the argument of logarithm. However, given $p_l\geq 0$, function $\log_2 \big( \frac{1+p_l \alpha_l}{1+p_l \beta_{l, j}} \big)$ is only negative when $\alpha_l\leq\beta_{l,j}$. Therefore, we can drop the non-linear operator $[\cdot]^+$ by taking the sum over set $\mathcal{L}_j\doteq \{l:\alpha_l \ge \beta_{l,j}\}$. 
Let us define $f_j(\bp, \bt) \doteq \sum_{l\in\mathcal{L}_j} t_l \log_2\left( \frac{1 + p_l \alpha_l}{1 + p_l \beta_{l,j}} \right)$, where $\bp = [p_1,\cdots, p_L]$ and $\bt = [t_1,\cdots, t_L]$. Therefore, the optimization problem in \eqref{P1_opt_prob} will be as follows
\begin{align} \label{P2_opt_prob}
    \textbf{P2:} \,\, &\underset{c, p_l \ge 0, t_l \ge 0, \forall l}{\text{max}} \,\,\,\,  c  \quad  \nonumber \\
    &\text{s.t. } \text{C}_0 : f_j(\bp, \bt) \ge c , \,\,\,\, \forall \bp_j \in \hat{\mathcal{P}} 
    \nonumber\\
    &\quad\,\,\, \text{C}_1 : \sum_{l=1}^{L} t_l \leq 1, \quad  \text{C}_2 : \sum_{l=1}^{L} p_l t_l \leq P_\text{tx},   
\end{align} 
where $c$ is an auxiliary variable. In the following, we consider the optimization problem \textbf{P2}.

We introduce variables \( \nu_j \ge 0 \), \( \lambda \ge 0 \), and \( \mu \ge 0 \), 
which are the Lagrange multipliers associated with constraints $\text{C}_0$, $\text{C}_1$, and $\text{C}_2$, respectively.
Now, we adopt a Lagrange dual formulation for \textbf{P2} in \eqref{P2_opt_prob}
\begin{align}
\mathcal{L}(\bp, \bt, c; \boldsymbol{\nu}, \lambda, \mu)
= &c + \sum_{j=1}^J \nu_j  \big( \sum_{l\in\mathcal{L}_j}t_l \log( \frac{1 + p_l \alpha_l}{1 + p_l \beta_{l,j}}  ) - c \big) \nonumber\\
&+ \lambda ( 1 - \sum_{l=1}^L t_l )
+ \mu ( P_\text{tx} - \sum_{l=1}^L p_l t_l ),
\end{align}
where $\boldsymbol{\nu}=[\nu_1,\dots,\nu_J]$. Note that we have changed $\log_2(\cdot)$ into the natural logarithm to simplify the following derivations. One can consider that the variables $\nu_j$ and $c$ are changed into $\nu_j/\log(2)$ and $c\log(2)$, respectively. Therefore, we can change $\log_2(\cdot)$ in $f_j(p,t)$ into $\log(\cdot)$.
By applying the Karush–Kuhn–Tucker (KKT) conditions, we derive analytical solutions for the primal variables $p_l$ and $t_l,\,\,\forall l$, as a function of the dual variables $\lambda$ and $\mu$. These conditions are given below.

\noindent
\textbf{Stationarity conditions:}
\begin{subequations}
\begin{align}
&\frac{\partial \mathcal{L}}{\partial c} = 1-\sum_{j=1}^J \nu_j= 0  \\
&\frac{\partial \mathcal{L}}{\partial t_l} = -\lambda - \mu p_l + \sum_{j:\,\, l\in\mathcal{L}_j} \nu_j \log \big( \frac{1 + p_l \alpha_l}{1 + p_l \beta_{l,j}} \big) = 0
\label{eq:Lagrange_D_t}\\
&\frac{\partial \mathcal{L}}{\partial p_l} = -\mu t_l + t_l \sum_{j:\,\, l\in\mathcal{L}_j}  \nu_j \big( \frac{\alpha_l}{1 + p_l \alpha_l} - \frac{\beta_{l,j}}{1 + p_l \beta_{l,j}} \big) = 0. 
\label{eq:Lagrange_D_p}
\end{align}
\end{subequations}

\noindent
\textbf{Primal feasibility conditions:} 
\begin{align}
\begin{cases}
    t_l \ge 0, \forall l\\
    p_l \ge 0, \forall l \\
    f_j(\bp, \bt) \ge c, \forall j \\
    \sum_{l=1}^L t_l \le 1 \\
    \sum_{l=1}^L p_l t_l \le P_\text{tx}.
\end{cases}
\end{align}

\noindent
\textbf{Dual feasibility conditions:} 
$\nu_j \ge 0, \forall j$, $\lambda \ge 0$, $\mu \ge 0$.

\noindent
\textbf{Complementary slackness conditions:} 
\begin{align}
\begin{cases}
    \nu_j (\sum_{l\in\mathcal{L}_j}  t_l \log( \!\frac{1 + p_l \alpha_l}{1 + p_l \beta_{l,j}} \! )\! - c) =0, \forall j \\
    \lambda \left( \sum_{l=1}^L t_l - 1 \right) = 0 \\
    \mu \left( \sum_{l=1}^L p_l t_l -  P_\text{tx} \right) = 0.
\end{cases}
\end{align}

Next, we develop an iterative algorithm, where at each iteration, we update the solution towards meeting the above KKT conditions. Let us assume first that the worst-case Eve's location is unique, i.e., $j^* = \arg\min_j f_j(\bp, \bt)$. 
Therefore, constraint $\text{C}_0$ is active only for $j=j^*$, which from the complementary slackness conditions,  we obtain $\nu_{j} =0, \forall j\neq j^*$ and from the stationarity conditions, we get $\nu_{j^*} =1$. Now, given the worst-case Eve's location, 
we adopt a gradient update for power and time variables at each iteration in order to move in the direction of satisfying the stationarity conditions:
\begin{subequations}
\begin{align}
&p_l \leftarrow [p_l + \eta_p (g_l- \mu)t_l]^+, \quad \forall l
\label{eq:solve_p_1}\\
&t_l \leftarrow [t_l + \eta_t (q_l - \lambda - \mu p_l)]^+, \quad \forall l
\label{eq:solve_t},
\end{align}
\end{subequations}
where $g_l \doteq \frac{\alpha_l}{1 + p_l \alpha_l} - \frac{\beta_{l,j^*}}{1 + p_l \beta_{l,j^*}}$, $q_l \doteq \log\left( \frac{1 + p_l \alpha_l}{1 + p_l \beta_{l,j^*}} \right)$, and  $\eta_t$ and $\eta_p$ are step sizes. Note that the stationarity condition in \eqref{eq:Lagrange_D_p} can be solved analytically for a known $j^*$; however, a gradual update as in \eqref{eq:solve_p_1} is preferred since a significant variation in the variables introduces numerical instability due to the underlying change in $j^*$ from one iteration to the next.

Finally, dual variables $\mu$ and $\lambda$ are updated as follows to meet the dual feasibility and complementary slackness conditions:
\begin{subequations}
\begin{align}
&\lambda \leftarrow [\lambda + \eta_\lambda r_1]^+, \label{eq:dual1}\\
&\mu \leftarrow [\mu + \eta_\mu r_2]^+,\label{eq:dual2}
\end{align}
\end{subequations}
where $r_1 = \sum_{l=1}^L t_l - 1 $ and $r_2 = \sum_{l=1}^L p_l t_l - P_\text{tx}$ are constraint residuals and $\eta_\lambda$ and $\eta_\mu$ are step sizes.  

The proposed iterative solution for \textbf{P2} is summarized in Algorithm~\ref{algorithm_1}.

\begin{algorithm}[t] 
\caption{Proposed power and time allocation algorithm}
\label{algorithm_1}
\begin{algorithmic}[1] 
\Statex \hspace{-0.4cm}\textbf{Initialization}
\State Initialize feasible primal variables $p_l$, $t_l$, dual variables $\lambda, \mu \ge 0$, step sizes $\eta_t$, $\eta_p$, $\eta_\lambda$, $\eta_\mu$, and solution tolerance~$\epsilon$.

\Statex \hspace{-0.4cm}\textbf{Iteration Steps}
\State {Compute active constraint:}
$j^* = \arg\min_j f_j(\bp, \bt)$.
\State Update primal variables \( p_l \) from \eqref{eq:solve_p_1} and \( t_l \) from \eqref{eq:solve_t}. 

\State {Update dual variables $\lambda$ from \eqref{eq:dual1} and $\mu$ from \eqref{eq:dual2}.}
\State Stop if improvement in $c=\min_j f_j(\bp, \bt)$ is below  $\epsilon$. 

\Statex \hspace{-0.4cm}\textbf{Output}
\State Optimal \( p_l^*, t_l^*, \lambda^*, \mu^* \), and \( c^* = f_{j^*}(\bp^*, \bt^* ) \).
\end{algorithmic}
\end{algorithm}

\section{Simulation Results}
As an initial investigation and to draw some insight, we consider a system with two ideal narrow beams ($L=2$, an LoS path and a 10 dB weaker non-LoS path), where the leakage from one beam to another is negligible, which is valid for large $N_t$ at mmWave bands. 
We consider the following normalized parameter values:  
$\alpha_1 = \beta_{1,1}= 2 \text{ W}^{-1}$, $\alpha_2 =\beta_{2,2}= 0.2 \text{ W}^{-1}$, $\beta_{1,2}= \beta_{2,1}= 0 \text{ W}^{-1}$, $P_\text{tx}=10$ W. 

\begin{figure}
    \centering
    \includegraphics[trim={2.4cm 16cm 4.7cm 2.1cm},clip,width=1\linewidth]{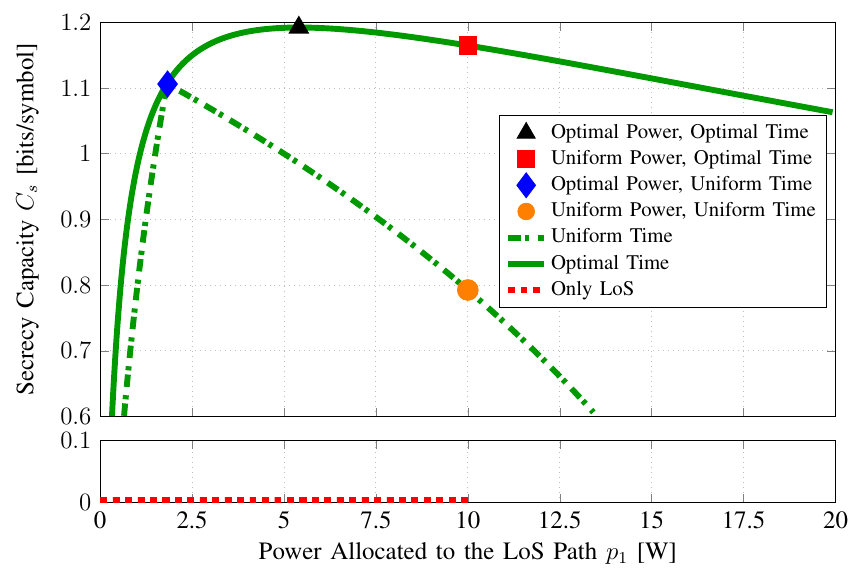} 
    \caption{Absolute secrecy capacity vs. the power allocated to the LoS path with the \textit{average} transmit power $P_\text{tx}=10$~W.}
    \label{fig_1}
\end{figure}

The secrecy capacity versus the power of the LoS path is plotted in Fig.~\ref{fig_1}. As observed, using only the LoS path for transmission is not secure, whereas joint coding even with uniform time and power allocation would consistently provide secrecy. Performance improves when only the powers are optimized, and improves even more when only the times are optimized \cite{ishtiaq2023beamsec, ishtiaq2025twc}. As shown, the scheme of joint power and time allocation outperforms the algorithms which only optimize either time or power. 
Moreover, Fig.~\ref{fig_1} shows that allocating less power to the LoS link than to the weaker non-LoS link maximizes secrecy rate.

\begin{figure}
    \centering
    \includegraphics[width=1\linewidth]{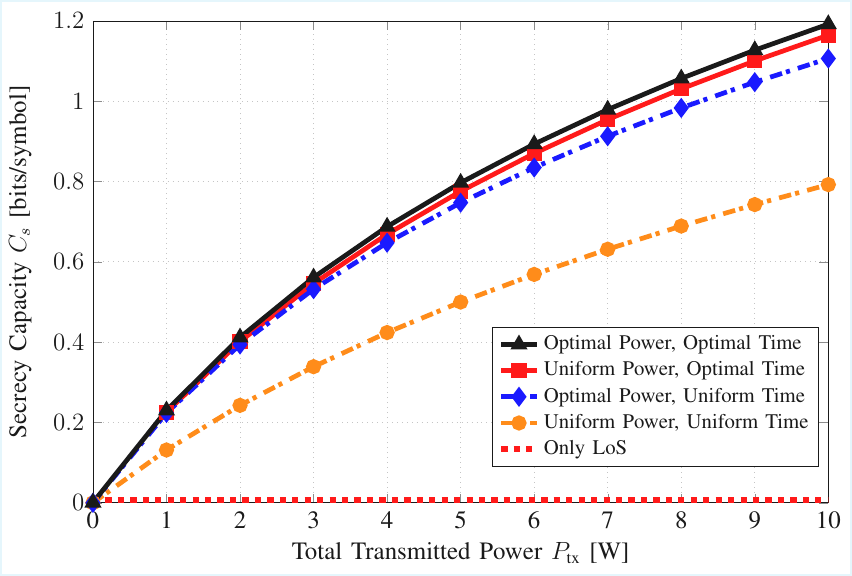} 
    \caption{Absolute secrecy capacity vs. the total transmitted power along the LoS and non-LoS paths.}
    \label{fig_2}
\end{figure}

The secrecy capacity versus the total transmitted power $P_\text{tx}$ (for both LoS and non-LoS paths) is plotted in Fig.~\ref{fig_2}. As observed, using only the LoS path does not provide secrecy. When both LoS and non-LoS paths are employed, the secrecy capacity is positive even with the simple scheme of uniform time and uniform power allocation. As shown, the scheme of joint power and time allocation outperforms the algorithms which only optimize either time (e.g., proposed in \cite{ishtiaq2023beamsec, ishtiaq2025twc}) or power. Note that Fig.~\ref{fig_1} is a special case of Fig.~\ref{fig_2} where $P_\text{tx}=10$ W, and confirms the secrecy capacities in Fig.~\ref{fig_2}.
Moreover, Fig.~\ref{fig_2} shows that the secrecy rate is an increasing function of the total transmitted power $P_\text{tx}$.
\section{Conclusion}
In this paper, we have employed CKM to enhance PLS in the presence of a passive Eve, without making any assumptions about Eve's location or CSI. The confidential message was transmitted using the highly directional mmWave band and was jointly encoded across multiple beams. We derived an algorithm for time and power allocation among the beams that maximizes the absolute secrecy rate under the worst-case scenario for Eve's location. The simulation results validate the performance of our scheme compared to previous methods, where either time or power is uniformly allocated, while the other is optimized. Furthermore, our results show that the secrecy capacity increases with the total transmitted power. 

\footnotesize 

\end{document}